\documentclass{iopjournal}
\usepackage[caption=false]{subfig}
\usepackage{cite}

\begin{document}

\articletype{Paper} 

\title{Anomalous diffusion in convergence to effective ergodicity}

\author{M. S\"uzen  \orcid{0000-0002-9460-7297}}

\affil{$^1$Member, American Physical Society, College Park, Maryland, United States} \\
\affil{$^2$Resident Researcher, Assia, CY 5561, Cyprus}

\email{mehmet.suzen@physics.org}


\begin{abstract}
The nature of diffusion is usually studied for particles or time-evolving systems. 
Similar in principle, such studies can be conducted by tracking how a given function 
of observable properties evolves over time—akin to the evolution of observable 
functions—referred to as {\it functional-diffusion}. This is not the same as 
the system's individual trajectories, but can be regarded as a meta-trajectory. 
Following this idea, we measure how the approach to ergodicity evolves over 
time for the observed magnetization of a full Ising model with an external field. 
We compute the diffusive behavior of the functional across a range of temperatures
via Metropolis and Glauber single-spin-flip dynamics. The system's ensemble-averaged 
dynamics are computed using expressions from the exact solution. Power-law behavior 
in the approach to ergodicity provides a classification of anomalies 
in {\it functional-diffusion}, demonstrating nonlinear anomalous behavior
over different temperature and field ranges. Studying the ergodicity convergence 
of these meta-trajectories can help validate and enhance the pedagogical 
understanding of nonequilibrium thermodynamic systems.
\end{abstract}

\section{Introduction}

Brownian motion is arguably one of the landmark concepts in statistical 
physics that attracted Einstein’s interest early on \cite{einstein1905}. 
Its importance in formulating statistical mechanics has been recently 
reviewed in 250th-anniversary publication \cite{dattagupta25}. A major 
observable in tracing Brownian motion is how the accumulated displacement 
curve behaves over time; specifically, a linear relationship without an 
intercept corresponds to {\it normal diffusion}. If the displacement 
curve of trajectories shows a power-law scaling over time other than 
this linear relationship, the behavior is called 
{\it anomalous diffusion} \cite{metzler98, metzler00, castiglione}. There 
is a recent surge of interest in using machine learning techniques to
analyze anomalous diffusion data \cite{munoz21, sposini22, seckler22, cai25}. 
For example, characterization using transformer architectures is quite 
novel given the growing interest in attention mechanisms within the 
machine learning community \cite{firbas23}.

Along these lines, the dynamics of cooperation among assemblies of 
independent units has been studied in this context \cite{wannier45a}, 
such as in the model of magnetic material \cite{ising25a, brush67a, baxter} 
and the state of a neuron \cite{little74, hopfield}. Measuring the diffusion
behavior in this discrete case would not only yield mathematically challenging 
consequences but also provide insights into characteristics of the Ising model. 
In our described scenario, this applies to the {\it functional} of trajectories. 
This distinction is the core driver in our work, where the functional—defined 
on the function that explains convergence to ergodicity for total magnetization
—is the diffusing process, rather than the trajectory of the discrete units of 
the Ising model.

Ergodicity is always preserved for the finite one-dimensional Ising model under 
periodic boundary conditions, but only after a sufficiently long time at finite 
temperatures \cite{fierro19}. Consequently, the question of how the system
approaches ergodicity during the initial time window manifests \cite{suezen2014a}. 
This remains an open research question due to the nonequilibrium nature of these 
regimes \cite{peliti21}. Such time-window dependence of ergodicity has been studied 
recently for self-gravitating systems \cite{souza23}, artificial spin 
ice \cite{saccone23, crater25a}, and laser speckle dynamics \cite{sdobnov}.
The observation of distinct non-ergodic regimes in the initial time window 
does not affect the established result that the one-dimensional Ising model remains 
ergodic and exhibits no phase transition for long times.

Convergence to ergodicity in this kind of cooperative dynamics has been explored 
and established \cite{suezen2014a} using the Thirumalai-Mountain (TM) fluctuation 
metric \cite{mountain89me, thirumalai1989ergodic}. Consequently, the power laws 
that emerge from the time evolution of the rate of effective ergodic convergence 
must be quantified. Understanding ergodicity under these circumstances is 
interesting not only for its fundamental importance in statistical 
mechanics \cite{dorfman99a} but also for its implications in real-world applications. 
These include understanding disruptions in neural networks for 
dementia \cite{thuraisingham2015a}, the realization of associative memory in 
solid-state devices \cite{hu2015associative}, the nature of economic 
utility \cite{peters2016evaluating}, and optical lattice 
dynamics \cite{schreiber2015a}.

The formulation of the Ising model, the concept of effective ergodicity 
and its measurement, along with an investigation of different power laws for 
functional-diffusion, are summarized in Section~\ref{sec:ergo}. Our extensive 
experimental results are presented in Section~\ref{sec:results}. The 
conclusion is given in Section~\ref{sec:sum}.

\section{Ergodicity convergence as diffusion phenomena} \label{sec:ergo}

We first introduce the Ising model as a lattice with \(N\) sites, labeled 
as \(\{s_{i}\}_{i=1}^{N}\), which can take two values, such as \(\{+1,-1\}\). 
These values imply spin up or down in the Ising model \cite{ising25a, brush67a, baxter} 
or activation in neuronal systems \cite{little74, hopfield}. The total energy, 
given by the Hamiltonian of the system, can be expressed via two interactions—one
due to nearest neighbors (NN) and one due to an external field—with 
coefficients \(J\) and \(H\), respectively:

 \begin{eqnarray}\label{eq:Ham1}
  \mathcal{H}({s_{i}}, J, H) = J \Big( (\sum_{i=1}^{N-1} s_{i} s_{i+1}) + (s_{1} s_{N}) \Big) + H \sum_{i=1}^{N} s_{i}.
\end{eqnarray} 

The term \(s_{1}s_{N}\) is imposed by the periodic boundary conditions, which provide 
translational invariance, making the model a closed chain of interacting units. The
thermal scale is expressed in terms of \(\beta =\frac{1}{k_{B}T}\), and the corresponding 
coefficients for NN and the external field are scaled by \(\beta \),

 \begin{eqnarray}
  J_{s} = \beta J, \qquad h=\beta H.
\end{eqnarray} 

Here, \(J_{s}\) represents the local interaction strength scaled by the thermal scale \(\beta \).
The analytic expression for the finite-size magnetization, \(M_{E}(N,\beta ,H)\), is obtained 
using the transfer matrix method. The time evolution of \(M_{E}(N,\beta ,H)\) is computed 
via a Monte Carlo procedure employing Metropolis and Glauber \textit{single-spin-flip dynamics}. 
Further details can be found in a previous work \cite{suezen2014a}, which focused solely on 
the underlying modified TM-metric for measuring ergodicity.

\begin{figure}[ptb]
  \centering
  \subfloat[\label{fig:KGlauberN1024beta1dot16}]{\includegraphics[width=0.5\columnwidth]{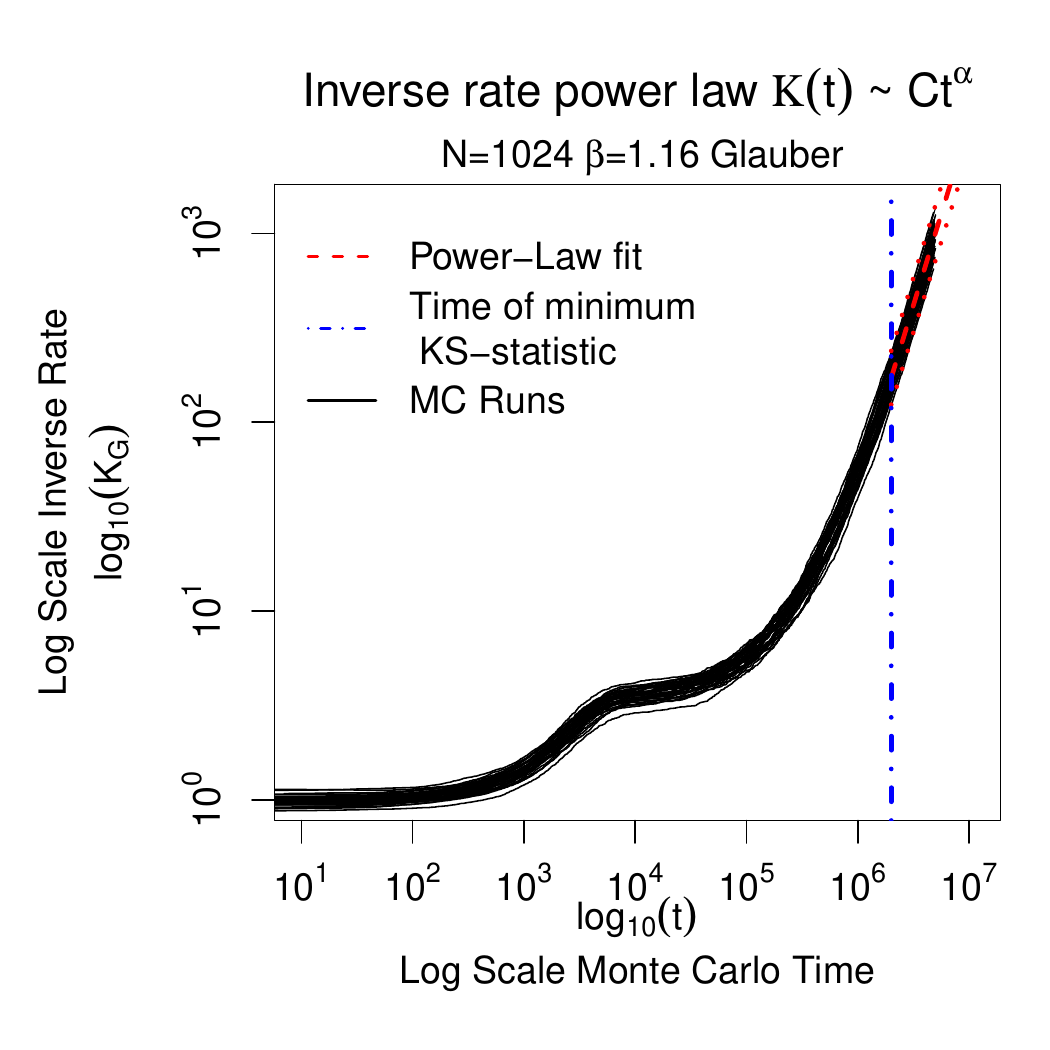}}
  \subfloat[\label{fig:KGlauberN1024field1dot4}]{\includegraphics[width=0.5\columnwidth]{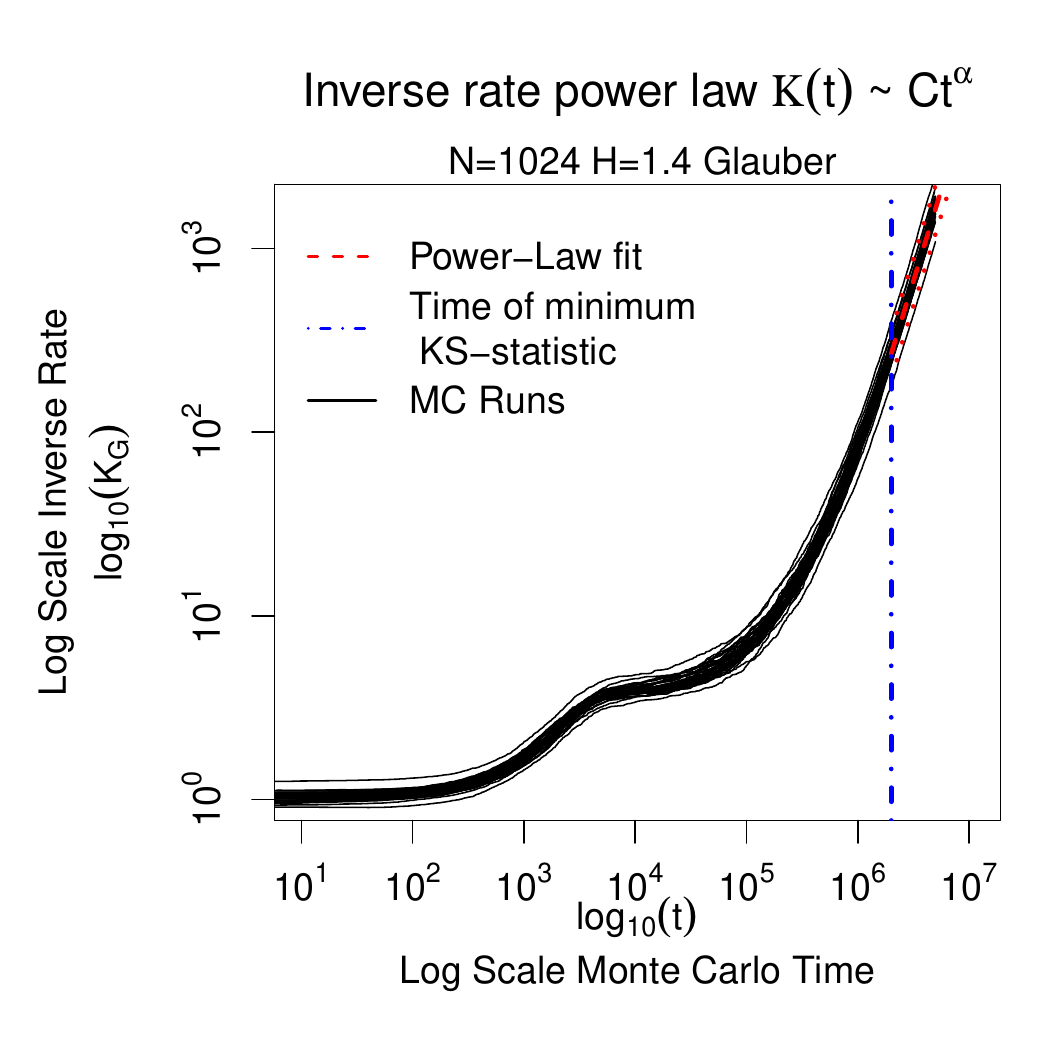}}
  \caption{Diagnostic plots of \(40\) different runs for \(N=1024\) with Glauber dynamics for the evolution 
           of \(K(t)\) log-log regression are shown. These include: (a) A field at \(H=1.0\) 
           and inverse temperature \(\beta =1.16\); (b) A field at \(H=1.4\) and inverse
           temperature \(\beta =1.0\). We identify the optimal time starting fit with a minimal 
           Kolmogorov-Smirnov statistic via a grid search. Uncertainties are processed after 
           independent runs. The computation of bias-corrected bootstrapping is applied to the 
           resulting scaling exponents \(\alpha \).}
\end{figure}

\begin{figure}[ptb]
  \centering
  \subfloat[\label{fig:powerLawsTimeGlauber}]{\includegraphics[width=0.5\columnwidth]{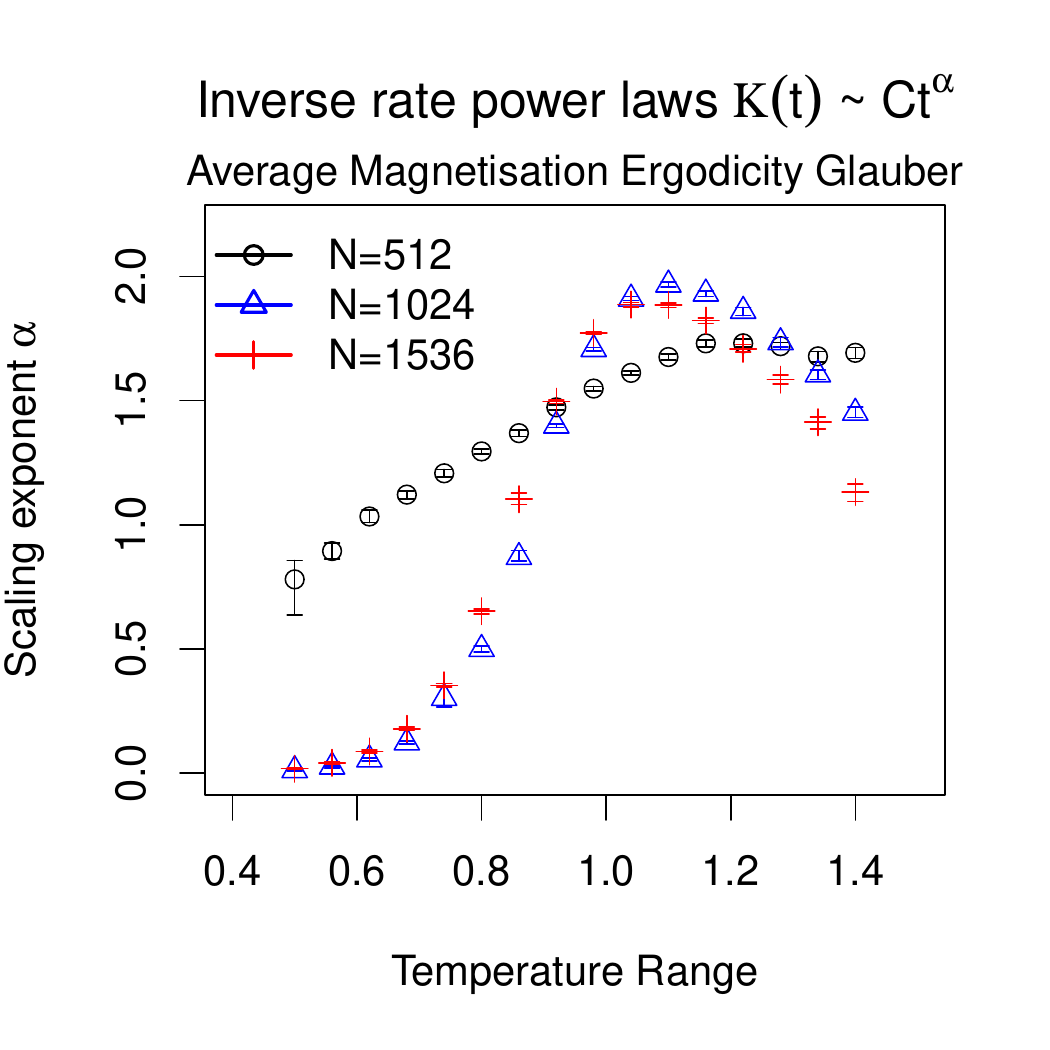}}
  \subfloat[\label{fig:powerLawsDistGlauber}]{\includegraphics[width=0.5\columnwidth]{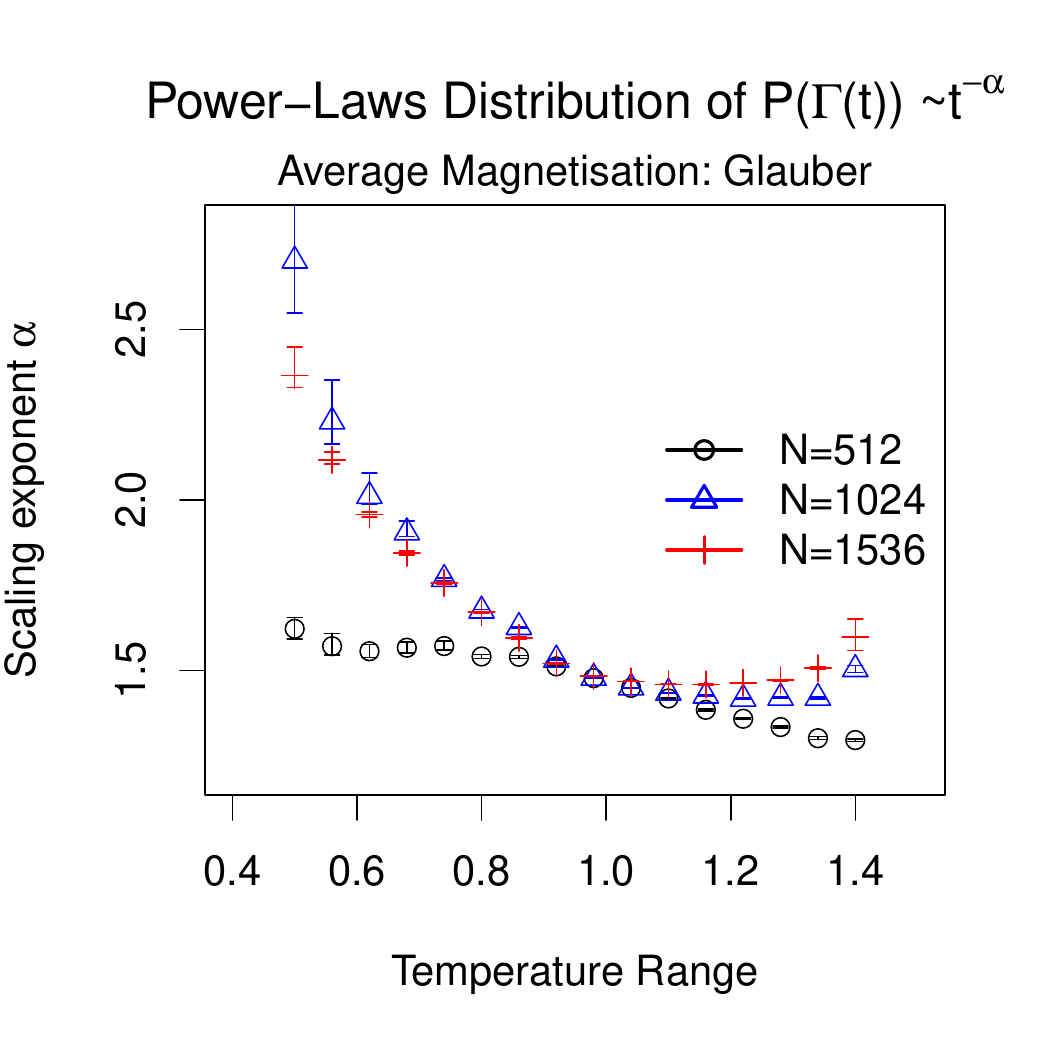}}
  \caption{Uncertainty is quantified for two power-law exponents \(\alpha \) over temperature 
          ranges:(a) the time evolution of \(K(t)\) via log-log regressions; 
          (b) the distribution of \(\Gamma (t)\) via analytical expressions from Newman 
          et al. Here, we plot the absolute value as \(\alpha \). In the fit, the functional 
          form involves a negative \(\alpha \). Uncertainties are processed after 
          independent runs, and the computation of bias-corrected bootstrapping is applied 
          to the resulting scaling exponents \(\alpha \).} 
\end{figure}

A type of ergodic dynamics suggested by Boltzmann is that trajectories of a many-body 
system will reach all regions of phase space in which the system is likely to be in 
thermodynamic equilibrium \cite{dorfman99a}. At this point, ensemble averages and time 
averages of the system produce nearly identical values. This implies that for a given 
observable \(g\) over a fixed region of phase space, the ensemble-averaged value can 
be recovered by time-averaged values. 

\begin{eqnarray}
  \label{eq:erEn}
  \langle g \rangle = lim_{t_{N} \to \infty} \int_{t_{0}}^{t_{N}} g(t) dt,
\end{eqnarray}

where \(\langle \rangle \) indicates the ensemble-averaged value. This basic 
definition is not standard in the literature \cite{mastatistical, mountain89me, kingman61}. 
Other forms of ergodicity demand that the system should visit all available phase-space 
partitions, which might not be possible for a finite physical system. Moreover,
the feasibility of this type of ergodicity is questioned \cite{gaveau2015ergodicity}. 
In practical terms, since partitions of the phase space are clustered around a few 
regions, effective ergodicity can be attained quickly \cite{mountain89me}.

On the other hand, {\it ergodicity} for {\it single-spin-flip dynamics}—essentially a Markov 
process—is defined by the accessibility of any given state point from another state point over 
time \cite{kingman61, pakes69}. In this sense, the Monte Carlo procedure used in this work is 
ergodic by construction for sufficiently long times.

{\it Effective ergodic convergence}, \(\Omega _{G}(t)\) for a given observable \(g\), can be 
constructed based on the fact that identical parts of cooperating units have identical 
characteristics in thermodynamic equilibrium \cite{mountain89me}. This is realized by the 
Thirumalai-Mountain (TM) \(G\)-fluctuation metric \cite{mountain89me, thirumalai1989ergodic}. 
Applications of ergodicity using the TM metric have appeared in recent studies, including 
biomolecular simulations \cite{grossfield09}, physical chemistry and machine learning 
landscapes \cite{wales10a}, seismology \cite{tiampo02a, tiampo03, tiampo07a, tiampo10a, 
cho10a, li12a}, neuromorphic computing \cite{baccetti24, barrows25}, and artificial 
spin ice \cite{saccone23, crater25a}.

TM metric at a given time $t_{k}$ reads:
\begin{eqnarray}
 \label{eq:Og}
 \Omega_{G}(t_{k}) = \frac{1}{N} \sum_{j=1}^{N} \big[ g_{j}(t_{k}) - \langle g(t_{k}) \rangle \big]^{2},
\end{eqnarray}

{\it Functional-diffusion} occurs during the development of a function of an observable over time,
$F[O(t)]$, measured as its displacement from the initial conditions. In this sense, 
the time-evolution of  $\Omega_{G}(t_{k})^{-1}$  can be considered a form of 
functional-diffusion. Here, the time-averaged quantity is expressed as 
$g_{j}(t_{k})$, and $\langle g(t_{k}) \rangle$ is defined as the instantaneous ensemble-averaged 
value over all units: 

\begin{eqnarray}
  \label{eq:gEns}
  g_{j}(t_{k}) & = & \frac{1}{k} \sum_{i=0}^{k} g_{j}(t_{i}), \\
  \langle g(t_{k}) \rangle & = & \frac{1}{N} \sum_{j=1}^{N} g_{j}(t_{k}).
\end{eqnarray} 

Consequently, {\it the rate of ergodic convergence} is measured as

\begin{eqnarray}
  \label{eq:Dg1}
  \Gamma(t) = \frac{\Omega_{G}(t)}{\Omega_{G}(0)} \to \frac{1}{t \cdot D_{G}},
\end{eqnarray}

with the diffusion coefficient $D_{G}$. A similar measure of ergodicity is used 
in simple liquids \cite{mountain89me, de2005diagnosing} and earthquake fault 
networks \cite{tiampo03a, tiampo2007a}.

{\it The inverse rate of ergodic convergence} is defined by $K(t)$:

\begin{eqnarray}
  \label{eq:DgKa}
   K(t) =  \frac{\Omega_{G}(0)}{\Omega_{G}(t)} \to t \cdot D_{G}. 
\end{eqnarray}

$\Omega_{G}$ is defined as a measure of ergodicity of the total magnetization in the Ising Model 
at time $t_{k}$, as a function of temperature and external field.

\begin{eqnarray}
  \label{eq:magG}
   \Omega_{M} (t_{k}, N, \beta, h)  & = & \big[ M_{T}(t_{k}) - M_{E}\big]^{2}, \\
                             M_{T}  & = & \frac{1}{k} \sum_{i=0}^{k} M(t_{i}), \nonumber \\
\end{eqnarray}

where $M_{T}(N, \beta, h)$ and $M_{E}( N, \beta, h)$ correspond to time-averaged and 
ensemble-averaged total magnetization, respectively. Since ensemble averages are computed 
analytically, our approach uses a modified TM-metric, leading to a more accurate 
assessment of the ensemble averaging. Note that the exact expression for $M_{E}$ is 
used \cite{suezen2014a}.

We aim to investigate the behavior of functional-diffusion. While the rate 
of ergodic convergence $\Gamma (t)$ and the inverse rate of ergodic 
convergence $K(t)$ are defined by the algebraic formulae used to compute 
them at a given time point, the manifestation of their diffusive behavior 
is not immediately obvious from these definitions. To quantify and explain 
the diffusive behavior of these two ergodic convergence functions, 
we generate dynamical data from the expressions and investigate 
the power-law behavior experimentally—specifically, by fitting the 
dynamical data. This is a phenomenological approach, rather than a 
formulation of the behavior via differential equations.

The first power law reads:

\begin{eqnarray}
  \label{eq:DgKaP2}
   K(t)  \to C \cdot t^{\alpha}. 
\end{eqnarray}

where $C$ is the generalized diffusion coefficient, $t$ is the Monte Carlo time, 
and $\alpha$  is a positive exponent. We refer to this as a time-dependent power law. 
We follow a phenomenological approach here, supported by strong theoretical foundations 
suggesting that such power laws emerge from diffusion equations—specifically 
Fokker-Planck equations \cite{risken} and  L\`evy flights; see \cite{metzler00} 
and the references therein.

The second type of power law we seek is described by the distribution of $\Gamma(t)$:

\begin{eqnarray}
  \label{eq:DgKaP}
   P(\Gamma(t))  \to \Gamma(t)^{-\alpha}.
\end{eqnarray}

This is identified as a distributional power law, which is related to L\`evy flights or 
jump distributions. This approach is again supported by a theoretical 
justification \cite{metzler00}, in which we consider cumulative jumps toward the 
approach to ergodicity. Both types of power laws are investigated within datasets 
generated using analytical expressions during dynamical simulations. 

Regarding functional-diffusion, the physical characterization from conventional 
diffusion studies—such as gradients and higher-order operators applied to the 
displacement function—would be defined identically once the functional is computed 
from an observable. This implies that the result of the functional $F[O(t)]$ at 
each time point gives rise to a function of time. $K(t)$ is produced in such a 
manner, and its gradients can be considered similarly. In this work, we took a more 
empirical approach rather than using conventional diffusion-based differential 
equations for $K(t)$. We have made a physically intuitive argument here, rather 
than a mathematically involved definition.

We will explicitly state which power laws we are working with to avoid confusing
the various \(\alpha \) values. Power laws in complex systems and the methods for 
computing them from empirical data have been studied in depth using techniques 
pioneered by Newman et al. \cite{newman2005power,clauset2009power}. We performed 
extensive bias-corrected bootstrapping to determine the uncertainties for scaling 
exponents, diffusion coefficients, fit diagnostics, KS-distances (Kolmogorov-Smirnov 
statistics), adjusted R-squared values, and autocorrelation times. The KS-distance 
was critical in determining the optimal starting point for performing log-log 
regressions; this was achieved in an automated manner by finding the minimum 
KS-statistic over a search grid of time values. Diagnostic visual inspections were 
conducted for every parameter, and these visualizations were developed and automated. 
Importantly, we aimed to use averaging over multiple trajectories following the 
fitting procedures. We averaged the results of the power-law fits from each 
individual trajectory to obtain a set of resulting parameters. We then applied 
bootstrapping to this set of parameters to determine the uncertainties, yielding 
asymmetric error bars.

We also computed autocorrelation times using time self-correlations
of averaged magnetization, via identification of the plateau of the correlation 
times $C(t)$ of averaged-magnetization,
$$C(t_{k}) = \frac{1}{t_{k}} \sum_{i=0}^{k}  M(t_{0}) \cdot M(t_{i})$$. 
We identify that relaxation changes in the regions where anomalous convergence appears.

A finite-size scaling (FSS) analysis could demonstrate the universality of the results, 
depending on the number of discrete units and the temperatures of the Ising model. 
This is a standard requirement in the numerical simulation of statistical 
systems \cite{privman} due to the presence of unavoidable finite-size effects. 
A prominent result of FSS in this context is known as data collapse \cite{bhattacharjee01}. 
For example, data collapse has been investigated recently for complex 
networks \cite{serafino21} and deep learning scaling laws \cite{qiu24, biroli24}, 
highlighting its importance in contemporary research.

After applying a parametrized scaling function—the so-called scaling ansatz—and 
given that the selection function is taken as is, we aimed to show that the results 
are not size-dependent; that is, the multiple curves resulting from different 
scales converge.

We made the following FSS assumptions for temperature and size dependence:
$$u(\beta) = (\beta-\beta_{c}) N^a$$, 

the scaling function $f(u)=A+B (1+exp(c \cdot u+d))^{-1}$,  

and the power-law exponents $\alpha$ with the critical temperature  $\beta_{c}$:

$$\alpha(T, N) = N^{-b} f(u)$$. 

We identified the FSS scaling exponents $a$, $b$ and coefficients $A$, $B$, $c$ and $d$, 
using nonlinear optimization (Nelder-Mead \cite{nelder, rlang}) for the time-dependent 
power laws over various temperatures. Physically, $a$ and $b$ are  dimensionless 
exponents explaining how $\alpha$  and the functional dependence on temperature $u$ 
scale with the system size $N$. The parameters $c$, $d$, $A$ and $B$ are similarly 
dimensionless parameters describing the functional dependence of 
temperature $u(\beta)$; this function is not unique and involves an 
empirical assumption, the so-called {\it ansatz} or given functional 
form \cite{privman, melchert09, bhattacharjee01}. A plot of of $u$ vs. $f(u)$  
should depict a data collapse, where all curves under different size conditions 
align.

\begin{figure}[ptb]
  \centering
  \subfloat[\label{fig:powerLawsTimeGlauberField}]{\includegraphics[width=0.5\columnwidth]{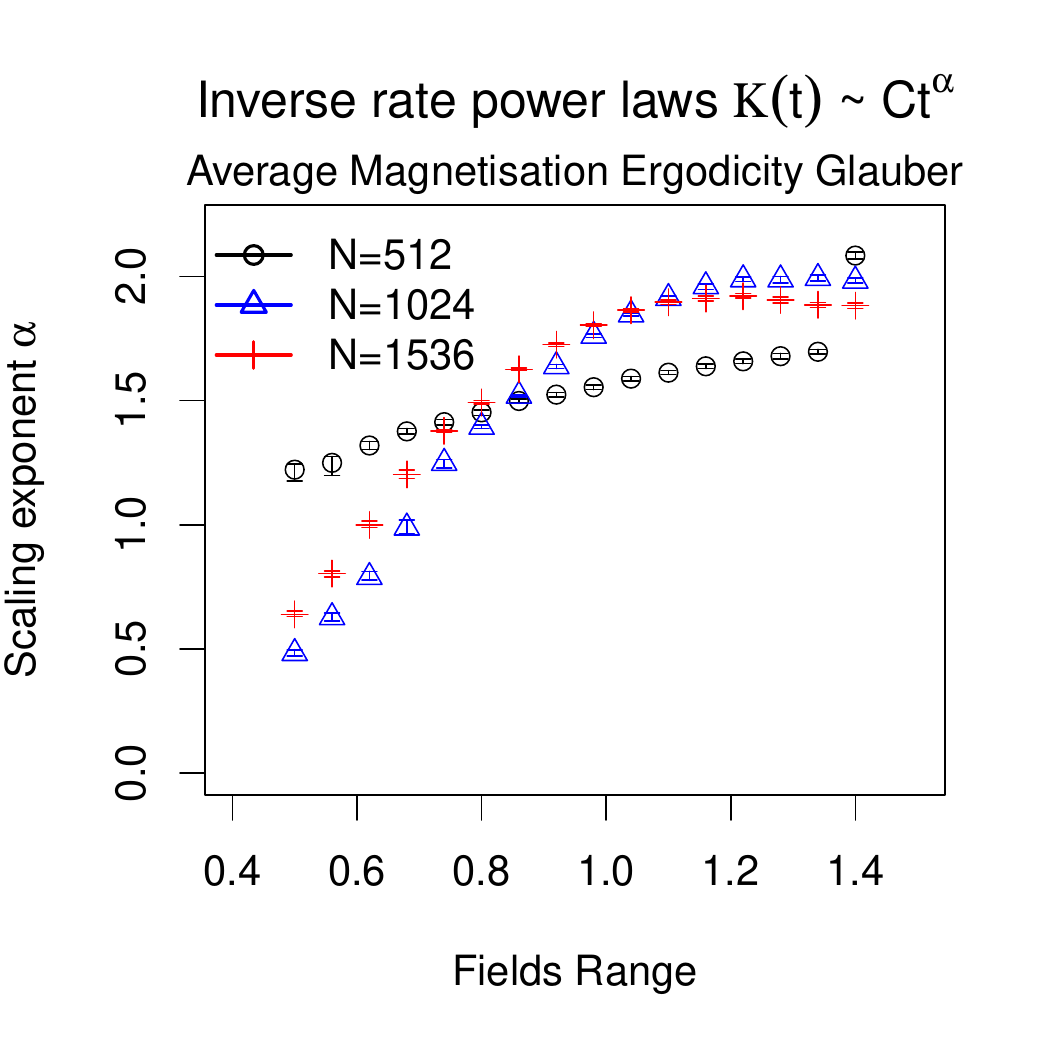}}
  \subfloat[\label{fig:powerLawsDistGlauberField}]{\includegraphics[width=0.5\columnwidth]{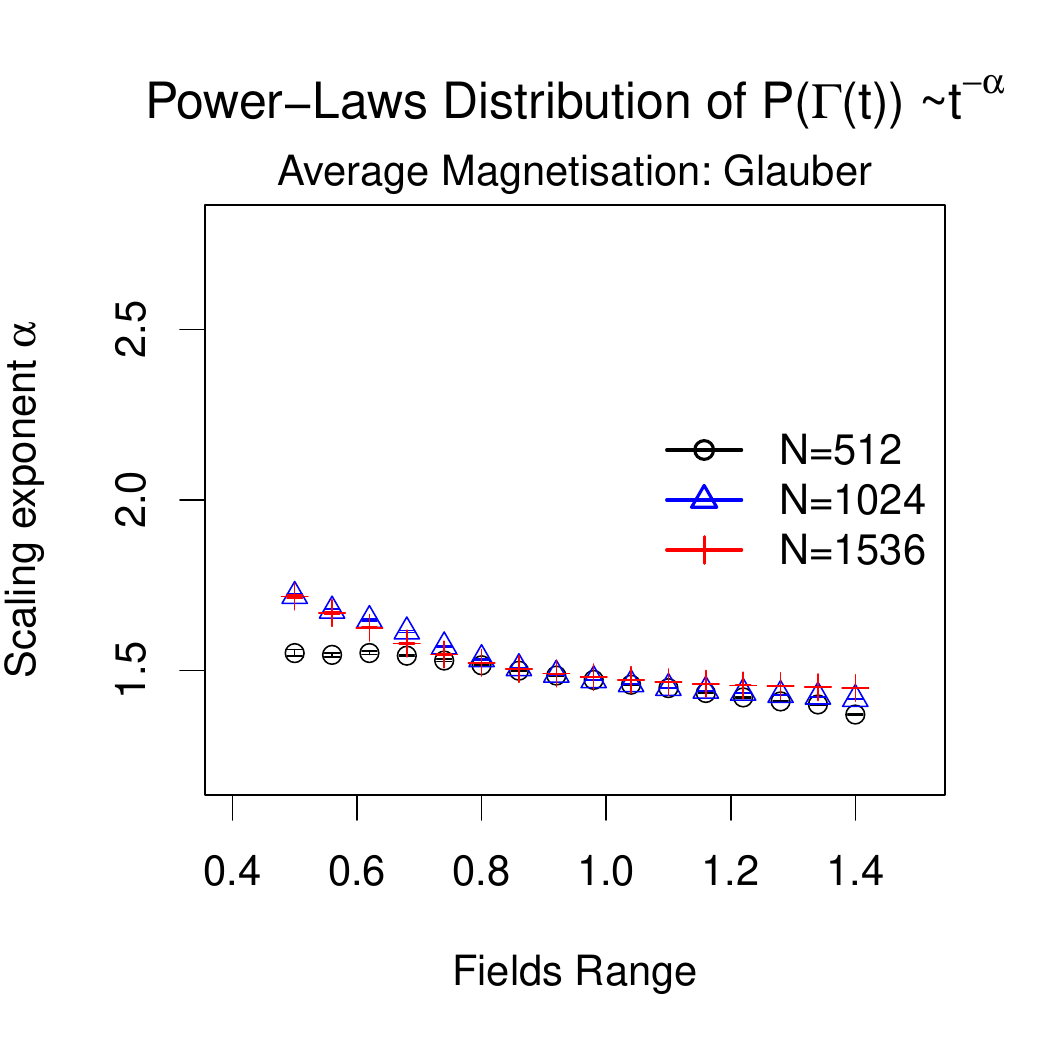}}
  \caption{Uncertainty is quantified for the two power-law $\alpha$ exponents over 
          field ranges at $\beta=1.0$: 
          (a) the time-evolution of $K(t)$,
          via log-log regressions;(b) the distribution of $\Gamma(t)$ 
          via analytical expressions from Newman et al. Here, we plot 
          the absolute value as $\alpha$. In the fit, the functional form involves 
          a negative $\alpha$.  Uncertainties are processed after independent runs, 
          and the computation of bias-corrected bootstrapping is applied to the 
          resulting scaling exponents $\alpha$.}
\end{figure}

\begin{figure}[ptb]
  \centering
  \subfloat[\label{fig:powerLawsTimeGlauberDiffusionCoeff}]{\includegraphics[width=0.5\columnwidth]{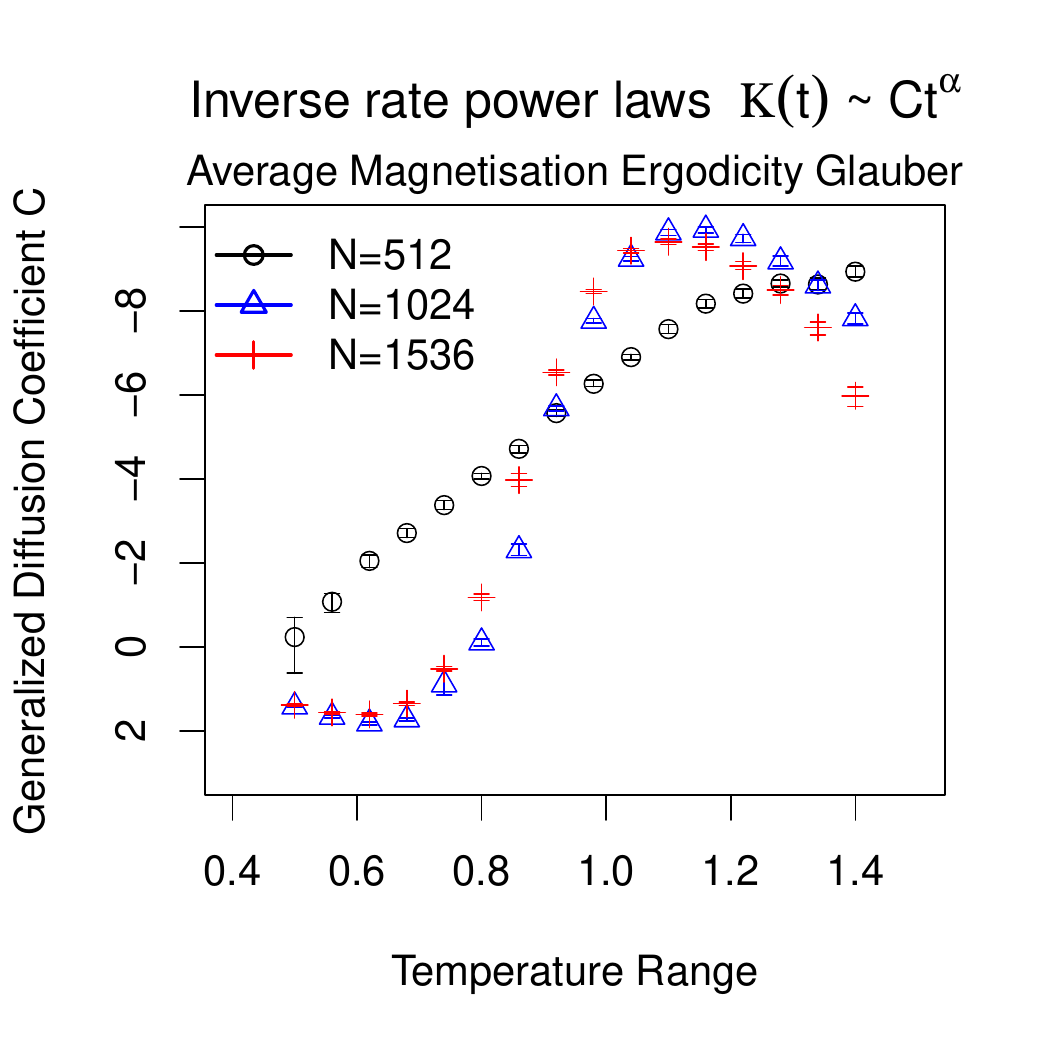}}
  \subfloat[\label{fig:powerLawsTimeGlauberDiffusionCoeffField}]{\includegraphics[width=0.5\columnwidth]{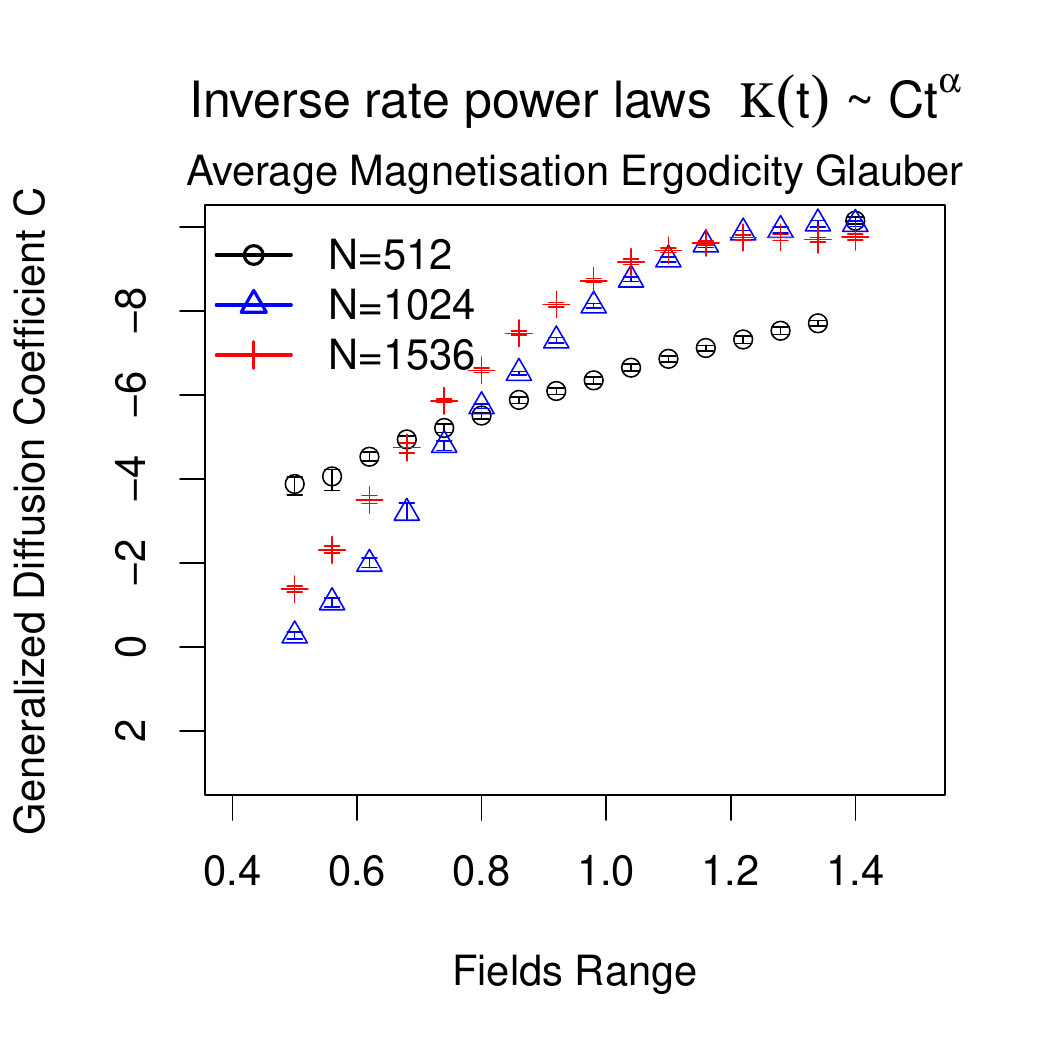}}
  \caption{Uncertainty is quantified for the power laws of the time evolution of 
           $K(t)$, using log-log regressions and diffusion coefficients: (a) Over 
           temperature ranges. (b) Over field ranges. Uncertainties are processed after 
           independent runs, and the computation of bias-corrected bootstrapping is 
           applied to the resulting diffusion coefficients.}
\end{figure}

\section{Experimental Results \label{sec:results}}

We perform extensive Metropolis and Glauber dynamics for the full Ising model 
and measure the rate of ergodic convergence, $\Gamma(t)$, and its inverse, $K(t)$.  
Here, we present the results of our findings. Our choice of parameter regions for 
the external field and temperature conforms to the principle of investigating regions 
around the canonical values $H=\beta=1.0$. The grid points were selected to balance 
the trade-off between our computational budget and a physically meaningful range.

The evolution of $K(t)$ for 40 independent runs is shown in 
Figures \ref{fig:KGlauberN1024beta1dot16} and \ref{fig:KGlauberN1024field1dot4} for 
different temperature and field conditions, along with diagnostics for the fitting 
starting points determined via optimal KS-distance analysis. These plots served as 
visual validation to ensure that each temperature, size, or field case was inspected, 
confirming that the fitted log-log regression indeed represents a power-law 
region within the simulated MC time, where we record $\Gamma(t)$ at Monte 
Carlo acceptance times. 

\begin{figure}[ptb]
  \centering
  \subfloat[\label{fig:acGlauberN1024beta0dot92}]{\includegraphics[width=0.5\columnwidth]{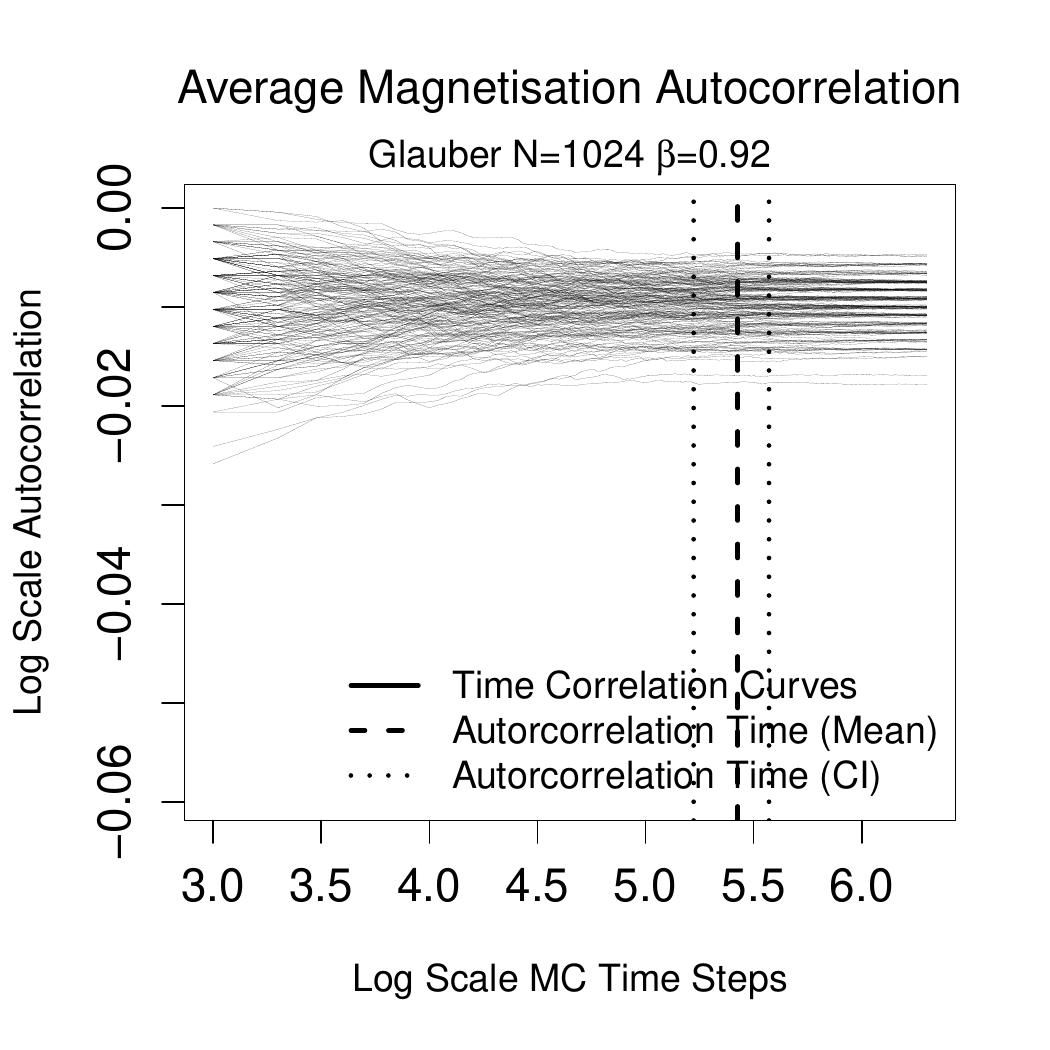}}
  \subfloat[\label{fig:acTimesGlauber}]{\includegraphics[width=0.5\columnwidth]{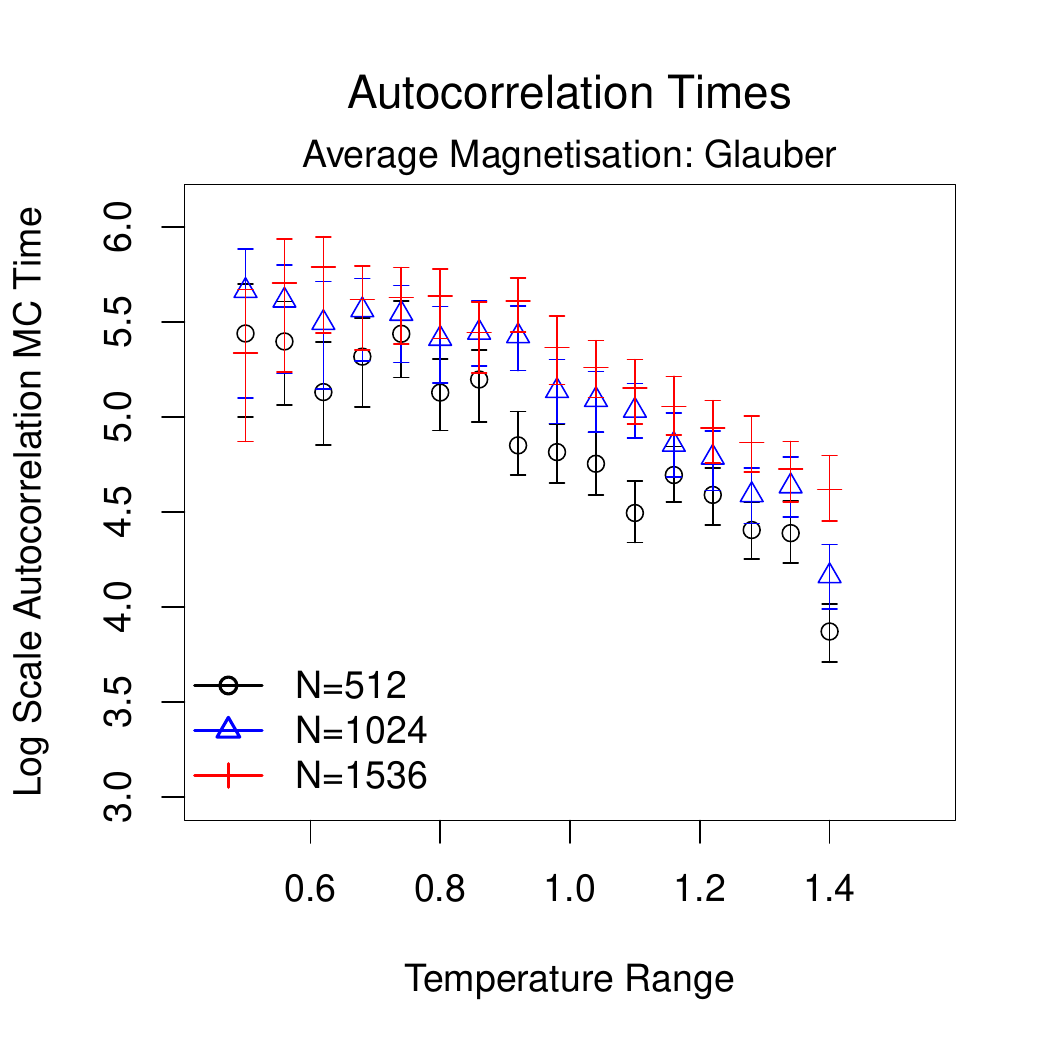}}
  \caption{The uncertainty of autocorrelation times was quantified using 200 independent runs.
           This involved:
           (a) a diagnostic plot for identifying the autocorrelation time;
           (b) a study of autocorrelation times across various temperature ranges and three 
           different system sizes. Uncertainties were processed after independent runs, and 
           the computation of bias-corrected bootstrapping was applied to the resulting 
           autocorrelation times.}
\end{figure}

\begin{table}[!tbp]
\label{tab:loglog}
\begin{center}
\begin{tabular}{rrrrrrr}
\hline\hline
\multicolumn{1}{c}{ikBT}&\multicolumn{1}{c}{alphaMean}&\multicolumn{1}{c}{alphaLower}&\multicolumn{1}{c}{alphaUpper}&\multicolumn{1}{c}{adjR2Mean}&\multicolumn{1}{c}{adjR2Lower}&\multicolumn{1}{c}{adjR2Upper}\tabularnewline
\hline
$0.50$&$0.0179$&$0.0145$&$0.0212$&$0.7133$&$0.6264$&$0.7830$\tabularnewline
$0.56$&$0.0416$&$0.0368$&$0.0448$&$0.8870$&$0.8179$&$0.9186$\tabularnewline
$0.62$&$0.0872$&$0.0818$&$0.0923$&$0.9710$&$0.9617$&$0.9769$\tabularnewline
$0.68$&$0.1784$&$0.1726$&$0.1852$&$0.9815$&$0.9775$&$0.9847$\tabularnewline
$0.74$&$0.3538$&$0.3462$&$0.3608$&$0.9897$&$0.9879$&$0.9913$\tabularnewline
$0.80$&$0.6524$&$0.6407$&$0.6617$&$0.9939$&$0.9934$&$0.9946$\tabularnewline
$0.86$&$1.1030$&$1.0812$&$1.1277$&$0.9969$&$0.9964$&$0.9972$\tabularnewline
$0.92$&$1.4959$&$1.4892$&$1.5030$&$0.9995$&$0.9994$&$0.9995$\tabularnewline
$0.98$&$1.7725$&$1.7665$&$1.7784$&$0.9999$&$0.9999$&$0.9999$\tabularnewline
$1.04$&$1.8869$&$1.8780$&$1.8947$&$0.9998$&$0.9997$&$0.9998$\tabularnewline
$1.10$&$1.8857$&$1.8754$&$1.8942$&$0.9995$&$0.9994$&$0.9996$\tabularnewline
$1.16$&$1.8223$&$1.8110$&$1.8334$&$0.9992$&$0.9990$&$0.9993$\tabularnewline
$1.22$&$1.7088$&$1.6951$&$1.7257$&$0.9990$&$0.9986$&$0.9991$\tabularnewline
$1.28$&$1.5846$&$1.5678$&$1.6033$&$0.9983$&$0.9979$&$0.9986$\tabularnewline
$1.34$&$1.4132$&$1.3864$&$1.4342$&$0.9966$&$0.9942$&$0.9975$\tabularnewline
$1.40$&$1.1320$&$1.0944$&$1.1654$&$0.9876$&$0.9832$&$0.9907$\tabularnewline
\hline
\end{tabular}\end{center}
\caption{ This table provides a diagnostic summary for the power-law fit of the 
 Glauber dynamics using log-log regression. Uncertainties are processed after
 independent runs, and the computation of bias-corrected bootstrapping is applied 
 to the resulting observables. Here, the upper and lower values represent uncertainties 
 around the mean, which appear in an asymmetric manner due to the bias-corrected bootstrapping procedure.
}
\end{table}

The power-law exponents $alpha$ for different sizes ($N=512, 1024, 1536$ ) are identified 
over inverse temperature ranges. The log-log regression results for time-dependent 
power laws—specifically \(K(t)\) observations over time—are shown in 
Figure \ref{fig:powerLawsTimeGlauber}. These are the functional-diffusion power laws, 
which align with conventional displacement power laws. We observe anomalous behavior 
in super-diffusive and sub-diffusive regions, with a normal diffusion region positioned 
in between. The power law depicted in Figure \ref{fig:powerLawsDistGlauber} has 
a different nature; it does not necessarily represent diffusive behavior but rather 
how the Monte Carlo dynamics evolve. From a simulation perspective, this captures 
the dynamical behavior resulting from Glauber single-spin-flip dynamics, validating 
that anomalies are present in the dynamics as well. More sophisticated cluster 
flips, such as Swendsen–Wang dynamics \cite{swendsen87}, may be required for regions 
with slow dynamics. Uncertainty estimates are obtained using bias-corrected 
bootstrapping to ensure the reliable identification of confidence 
intervals \cite{efron94, davison97}. We present a set of uncertainty quantifications 
for the Glauber dynamics \(K(t)\) power-law log-log plots in Table \ref{tab:loglog}. 
Similar observations have been made for the exponents $alpha$ in the field
variations shown in Figures \ref{fig:powerLawsTimeGlauberField} 
and \ref{fig:powerLawsDistGlauberField}. Generalized diffusion coefficients 
over time validate these findings, as seen in 
Figures \ref{fig:powerLawsTimeGlauberDiffusionCoeff} 
and \ref{fig:powerLawsTimeGlauberDiffusionCoeffField}. Finally, 
the identification of autocorrelation times from the time correlations
of the average magnetization indicates that different relaxation times 
match the observed diffusion regimes, as shown in
Figures \ref{fig:acGlauberN1024beta0dot92} and \ref{fig:acTimesGlauber}. 

\begin{table}[!tbp]
\begin{center}
\begin{tabular}{llrrrl}
\hline\hline
\multicolumn{1}{l}{df}&\multicolumn{1}{c}{type\_power\_law}&\multicolumn{1}{c}{N}&\multicolumn{1}{c}{D}&\multicolumn{1}{c}{pvalue}&\multicolumn{1}{c}{parameter}\tabularnewline
\hline
1&Distribution Temperature Range&$ 512$&$0.1875$&$0.9522$&alpha scale\tabularnewline
2&Distribution Temperature Range&$1024$&$0.1875$&$0.9522$&alpha scale\tabularnewline
3&Distribution Temperature Range&$1536$&$0.1250$&$0.9998$&alpha scale\tabularnewline
4&Distribution Field Range&$ 512$&$0.2500$&$0.7164$&alpha scale\tabularnewline
5&Distribution Field Range&$1024$&$0.0625$&$1.0000$&alpha scale\tabularnewline
6&Distribution Field Range&$1536$&$0.1250$&$0.9998$&alpha scale\tabularnewline
7&Time Temperature Range&$ 512$&$0.0625$&$1.0000$&alpha scale\tabularnewline
8&Time Temperature  Range&$1024$&$0.0625$&$1.0000$&alpha scale\tabularnewline
9&Timer Temperature Range&$1536$&$0.1250$&$0.9998$&alpha scale\tabularnewline
10&Time Temperature Range&$ 512$&$0.1250$&$0.9998$&C diffusion\tabularnewline
11&Time Temperature  Range&$1024$&$0.1250$&$0.9998$&C diffusion\tabularnewline
12&Timer Temperature Range&$1536$&$0.1250$&$0.9998$&C diffusion\tabularnewline
13&Time Field Range&$ 512$&$0.0625$&$1.0000$&alpha scale\tabularnewline
14&Time Field  Range&$1024$&$0.1875$&$0.9522$&alpha scale\tabularnewline
15&Time Field Range&$1536$&$0.1250$&$0.9998$&alpha scale\tabularnewline
16&Time Field Range&$ 512$&$0.0625$&$1.0000$&C diffusion\tabularnewline
17&Time Field  Range&$1024$&$0.1250$&$0.9998$&C diffusion\tabularnewline
18&Time Field Range&$1536$&$0.1875$&$0.9522$&C diffusion\tabularnewline
\hline
\end{tabular}\end{center}
\caption{ This diagnostic summary compares power laws across different 
ranges for both Glauber and Metropolis dynamics using two-sided 
KS tests. It includes mean $K(t)$ power laws and distributional 
mean $P(\Gamma(t))$ power laws. Uncertainties are processed after 
independent runs, and the computation of bias-corrected bootstrapping 
is applied to the resulting observables.}
\label{tab:glaubermetro}
\end{table}

\begin{figure}[ptb]
  \centering
  \subfloat[\label{fig:acGlauberN1024beta0dot92}]{\includegraphics[width=0.5\columnwidth]{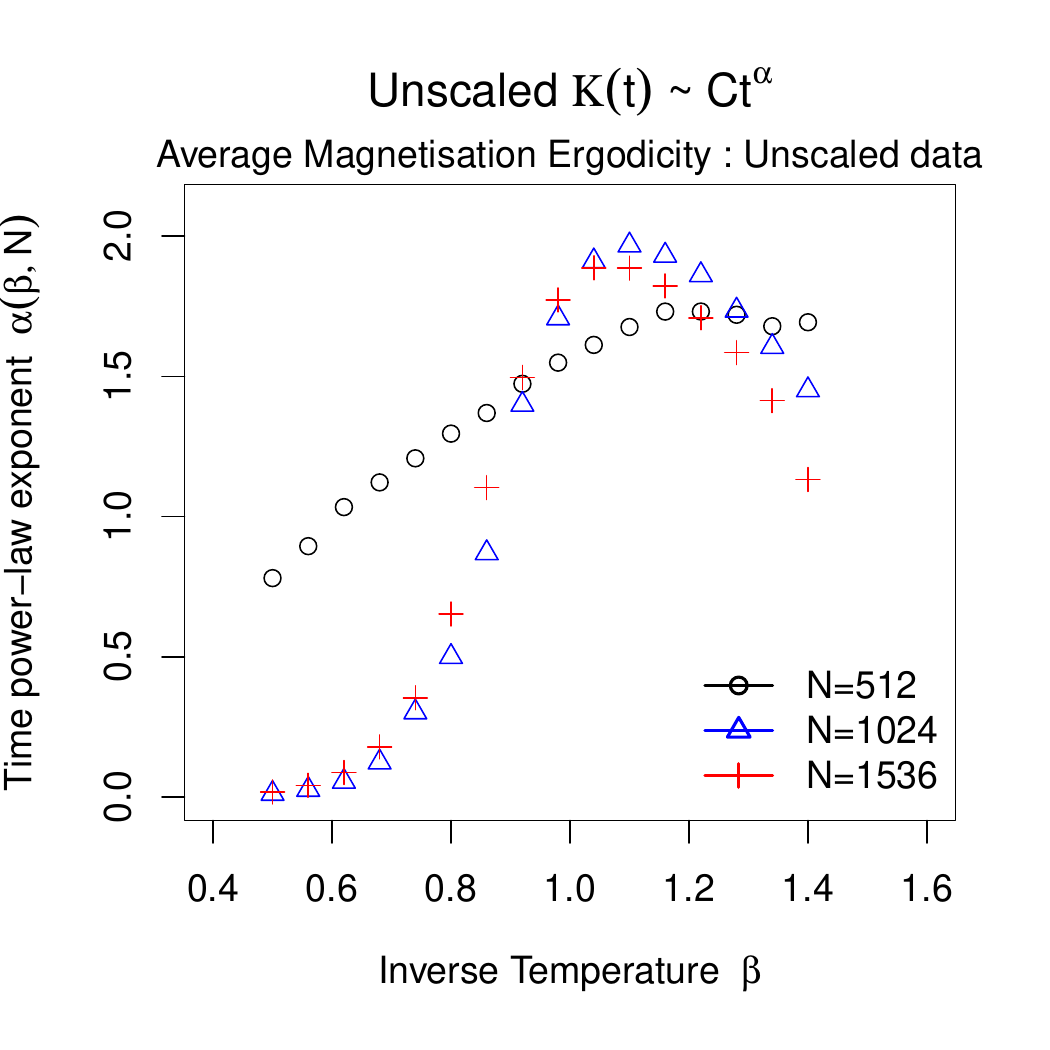}}
  \subfloat[\label{fig:acTimesGlauber}]{\includegraphics[width=0.5\columnwidth]{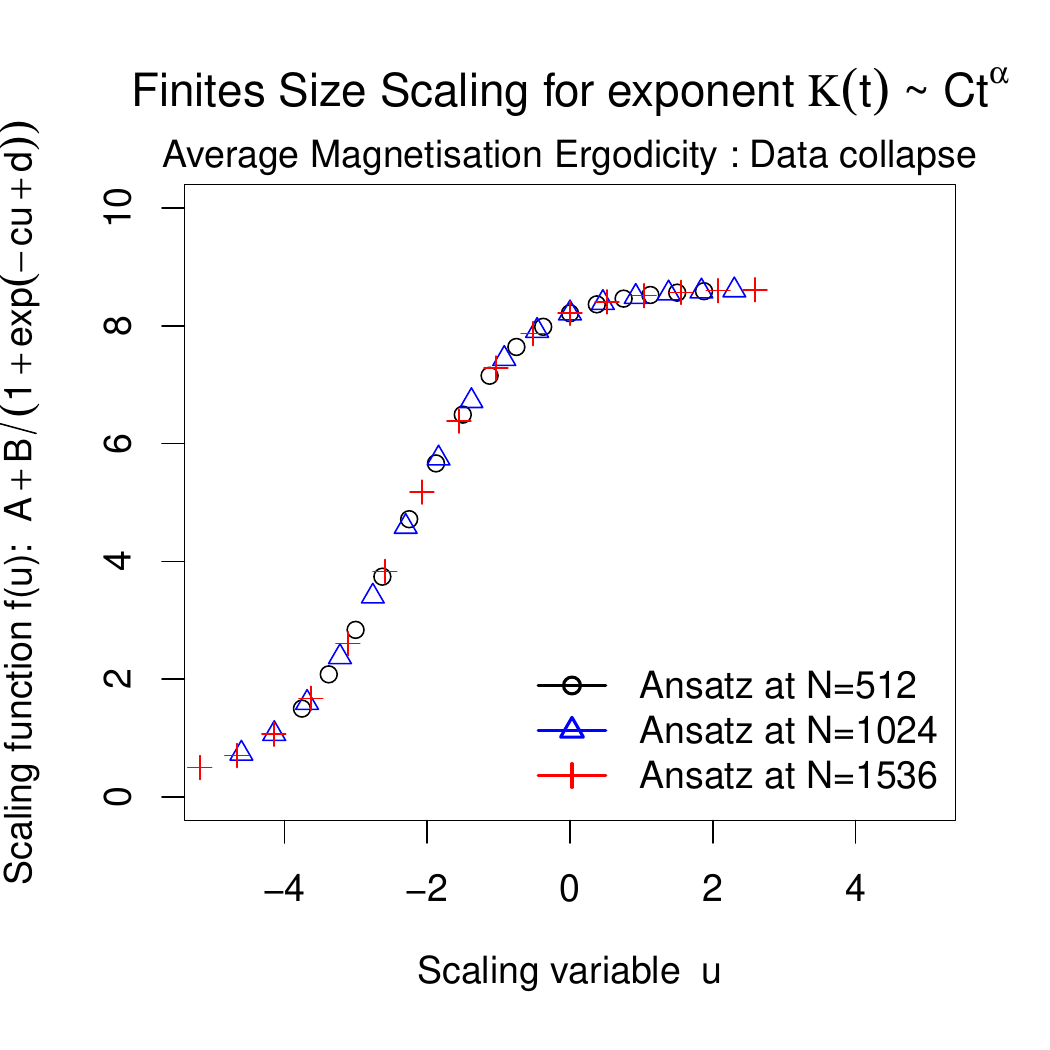}}
  \caption{These plots provide a demonstration of data collapse for the
           time-dependent power laws:
           (a) Unscaled exponents are plotted over the inverse temperature.
           (b) Data collapse is shown after applying the finite-size scaling analysis.
           }
  \label{figfss}
\end{figure}

Finite-size scaling (FSS) parameters are also computed using nonlinear optimization. 
The following FSS parameters were obtained:
$a=0.294$,  $b=0.2318$, $c=1.2526$, $d=-2.9509$, $A=0.2647$ 
and $B=8.3646 $. Nonlinear optimization of the FSS parameters was 
performed across different temperatures and sizes. We found 
that results exhibiting a so-called {\it data collapse} \cite{bhattacharjee01}.
Using the optimized parameters, {\it the data collapse} is shown in 
Figure \ref{figfss} via the scaling ansatz. This data collapse behavior 
independently validates our power-law results.

It is also established in Table \ref{tab:glaubermetro} that the results are
statistically significant for both Metropolis and Glauber dynamics.

In different parameter settings, the combination of spin-interaction and external 
field values drives the system into nonlinear regimes with varying correlation 
lengths, as demonstrated via self-autocorrelation functions. The approach to 
equilibrium over these ranges indeed demonstrates nonlinear relationships. 
Consequently, anomalous diffusion in the approach to ergodicity is an expected 
behavior rather than normal diffusive behavior over time.

\section{Conclusion\label{sec:sum}}

A new concept of {\it functional-diffusion} is introduced via a canonical example 
using the Ising model: the approach to ergodicity for total magnetization. 
Superdiffusive regimes are identified across different temperature and field 
ranges through the computation of power laws for the forward and inverse 
fluctuating metrics for ergodicity. These comprehensive results demonstrate, 
for the first time, a quantitative measure of anomalous convergence to 
ergodicity using power laws. This concept provides a pedagogical test bed 
for extending diffusive behavior beyond particle trajectories—specifically 
in the context of {\it the functional diffusion for the full Ising model}—and 
increases awareness of the potential for anomalous convergence in metric 
functionals, such as the TM metric.

Understanding of nonequilibrium and stochastic thermodynamic 
systems \cite{dorfman99a, peliti21} could be advanced by this concept 
and the study of their ergodic convergence, which illuminates the 
underlying statistical mechanics behavior.

As a limitation of this work, the term {\it functional-diffusion} remains a 
phenomenological analysis tool tied to describing the dynamics of the 
TM ergodicity measure in the one-dimensional Ising model. It requires 
more formalized and in-depth future study to become a stand-alone concept 
beyond the TM metric.

\section*{Data Availability}

The main diagnostic tables, as well as all data generation and 
analysis R notebooks \cite{rlang, chambers08}, are available in 
the public domain via a Zenodo repository \cite{suzen25zenV3} under 
an open-source license. The {\it IsingLenzMC} package was utilized 
for core data generation \cite{isinglenzmc}.

\section*{Acknowledgements}

We are grateful to Y. S\"uzen for her kind support and encouragement. 
The author would like to express gratitude to the referees for their 
constructive recommendations and for pointing out relevant literature, 
which made the computational work more robust.

\bibliographystyle{iopart-num}
\bibliography{suzen}

@article{einstein1905,
  title         = {{\"U}ber die von der molekularkinetischen Theorie der W{\"a}rme geforderte Bewegung von in ruhenden Fl{\"u}ssigkeiten suspendierten Teilchen},
  author        = {Einstein, Albert},
  journal       = {Annalen der physik},
  volume        = {322},
  number        = {8},
  pages         = {549--560},
  year          = {1905},
  publisher     = {Wiley Online Library},
  doi           = {10.1002/andp.19053220806},
  url           = {https://doi.org/10.1002/andp.19053220806}
}

@article{dattagupta25,
  author        = {Dattagupta, Sushanta and Ghosh, Aritra},
  title         = {Brownian-motion approach to statistical mechanics: Langevin equations, fluctuations, and timescales},
  journal       = {Physics of Fluids},
  volume        = {37},
  number        = {2},
  pages         = {027199},
  year          = {2025},
  month         = {02},
  issn          = {1070-6631},
  doi           = {10.1063/5.0255687},
  url           = {https://doi.org/10.1063/5.0255687}
}

@article{wannier45a,
  title         = {The Statistical Problem in Cooperative Phenomena},
  author        = {Wannier, G. H.},
  journal       = {Rev. Mod. Phys.},
  volume        = {17},
  issue         = {1},
  pages         = {50--60},
  numpages      = {0},
  year          = {1945},
  month         = {Jan},
  publisher     = {American Physical Society},
  doi           = {10.1103/RevModPhys.17.50},
  url           = {http://link.aps.org/doi/10.1103/RevModPhys.17.50}
}

@article{ising25a,
  title         = {Beitrag zur Theorie des Ferromagnetismus},
  author        = {Ising, Ernst},
  journal       = {Zeitschrift f{\"u}r Physik A Hadrons and Nuclei},
  volume        = {31},
  number        = {1},
  pages         = {253--258},
  year          = {1925},
  publisher     = {Springer},
  doi           = {10.1007/BF02980577},
  url           = {https://doi.org/10.1007/BF02980577}
}

@article{brush67a,
  title         = {History of the Lenz-Ising model},
  author        = {Brush, Stephen G},
  journal       = {Reviews of Modern Physics},
  volume        = {39},
  number        = {4},
  pages         = {883},
  year          = {1967},
  publisher     = {APS},
  doi           = {10.1103/RevModPhys.39.883},
  url           = {https://doi.org/10.1103/RevModPhys.39.883}
}

@book{baxter,
  title         = {Exactly solvable models in statistical mechanics},
  author        = {Baxter, RJ},
  year          = {1985},
  publisher     = {World Scientific},
  doi           = {10.1142/9789814415255_0002},
  url           = {https://doi.org/10.1142/9789814415255_0002}
}

@article{little74,
  title         = {The existence of persistent states in the brain},
  journal       = {Mathematical Biosciences},
  volume        = {19},
  number        = {1},
  pages         = {101--120},
  year          = {1974},
  doi           = {10.1016/0025-5564(74)90031-5},
  url           = {https://www.sciencedirect.com/science/article/pii/0025556474900315},
  author        = {W.A. Little}
}

@article{hopfield,
  author        = {J J Hopfield},
  title         = {Neural networks and physical systems with emergent collective computational abilities.},
  journal       = {Proceedings of the National Academy of Sciences},
  volume        = {79},
  number        = {8},
  pages         = {2554--2558},
  year          = {1982},
  doi           = {10.1073/pnas.79.8.2554},
  url           = {https://www.pnas.org/doi/abs/10.1073/pnas.79.8.2554}
}

@article{castiglione,
  title         = {On strong anomalous diffusion},
  journal       = {Physica D: Nonlinear Phenomena},
  volume        = {134},
  number        = {1},
  pages         = {75--93},
  year          = {1999},
  doi           = {10.1016/S0167-2789(99)00031-7},
  url           = {https://www.sciencedirect.com/science/article/pii/S0167278999000317},
  author        = {P. Castiglione and A. Mazzino and P. Muratore-Ginanneschi and A. Vulpiani}
}

@article{suezen2014a,
  title         = {Effective ergodicity in single-spin-flip dynamics},
  author        = {S{\"u}zen, M.},
  journal       = {Physical Review E},
  volume        = {90},
  number        = {3},
  pages         = {032141},
  year          = {2014},
  publisher     = {APS},
  doi           = {10.1103/PhysRevE.90.032141},
  url           = {https://doi.org/10.1103/PhysRevE.90.032141}
}

@book{dorfman99a,
  title         = {An introduction to chaos in nonequilibrium statistical mechanics},
  author        = {Dorfman, Jay Robert},
  number        = {14},
  year          = {1999},
  publisher     = {Cambridge University Press},
  doi           = {10.1017/CBO9780511628870},
  url           = {https://doi.org/10.1017/CBO9780511628870}
}

@book{peliti21,
  title         = {Stochastic thermodynamics: an introduction},
  author        = {Peliti, Luca and Pigolotti, Simone},
  year          = {2021},
  publisher     = {Princeton University Press},
  url           = {https://press.princeton.edu/books/hardcover/9780691201771/stochastic-thermodynamics}
}

@article{mountain89me,
  title         = {Measures of effective ergodic convergence in liquids},
  author        = {Mountain, Raymond D and Thirumalai, D},
  journal       = {The Journal of Physical Chemistry},
  volume        = {93},
  number        = {19},
  pages         = {6975--6979},
  year          = {1989},
  publisher     = {ACS Publications},
  doi           = {10.1021/j100356a019},
  url           = {https://doi.org/10.1021/j100356a019}
}

@article{thirumalai1989ergodic,
  title         = {Ergodic behavior in supercooled liquids and in glasses},
  author        = {Thirumalai, D. and Mountain, R.D. and Kirkpatrick, T.R.},
  journal       = {Physical Review A},
  volume        = {39},
  number        = {7},
  pages         = {3563},
  year          = {1989},
  publisher     = {APS},
  doi           = {10.1103/PhysRevA.39.3563},
  url           = {https://doi.org/10.1103/PhysRevA.39.3563}
}

@article{de2005diagnosing,
  title         = {Diagnosing broken ergodicity using an energy fluctuation metric},
  author        = {de Souza, Vanessa K and Wales, David J},
  journal       = {The Journal of Chemical Physics},
  volume        = {123},
  number        = {13},
  pages         = {134504},
  year          = {2005},
  publisher     = {AIP Publishing},
  doi           = {10.1063/1.2035080},
  url           = {https://doi.org/10.1063/1.2035080}
}

@article{tiampo03a,
  title         = {Ergodic dynamics in a natural threshold system},
  author        = {Tiampo, KF and Rundle, JB and Klein, W and Martins, JS S\'{a} and Ferguson, CD},
  journal       = {Physical Review Letters},
  volume        = {91},
  number        = {23},
  pages         = {238501--238501},
  year          = {2003},
  publisher     = {American Physical Society},
  doi           = {10.1103/PhysRevLett.91.238501},
  url           = {https://doi.org/10.1103/PhysRevLett.91.238501}
}

@article{tiampo2007a,
  title         = {Ergodicity in natural earthquake fault networks},
  author        = {Tiampo, KF and Rundle, JB and Klein, W and Holliday, J and Martins, JS S\'{a} and Ferguson, CD},
  journal       = {Physical Review E},
  volume        = {75},
  number        = {6},
  pages         = {066107},
  year          = {2007},
  publisher     = {APS},
  doi           = {10.1103/PhysRevE.75.066107},
  url           = {https://doi.org/10.1103/PhysRevE.75.066107}
}

@article{schreiber2015a,
  title         = {Observation of many-body localization of interacting fermions in a quasirandom optical lattice},
  author        = {Schreiber, Michael and Hodgman, Sean S and Bordia, Pranjal and L\"{u}schen, Henrik P and Fischer, Mark H and Vosk, Ronen and Altman, Ehud and Schneider, Ulrich and Bloch, Immanuel},
  journal       = {Science},
  volume        = {349},
  number        = {6250},
  pages         = {842--845},
  year          = {2015},
  publisher     = {American Association for the Advancement of Science},
  doi           = {10.1126/science.aaa7432},
  url           = {https://doi.org/10.1126/science.aaa7432}
}

@article{gaveau2015ergodicity,
  title         = {Is ergodicity a reasonable hypothesis for macroscopic systems?},
  author        = {Gaveau, Bernard and Schulman, Lawrence S},
  journal       = {The European Physical Journal Special Topics},
  volume        = {224},
  number        = {5},
  pages         = {891--904},
  year          = {2015},
  publisher     = {Springer},
  doi           = {10.1140/epjst/e2015-02434-7},
  url           = {https://doi.org/10.1140/epjst/e2015-02434-7}
}

@article{newman2005power,
  title         = {Power laws, Pareto distributions and Zipf's law},
  author        = {Newman, Mark EJ},
  journal       = {Contemporary Physics},
  volume        = {46},
  number        = {5},
  pages         = {323--351},
  year          = {2005},
  publisher     = {Taylor \& Francis},
  doi           = {10.1080/00107510500052444},
  url           = {https://doi.org/10.1080/00107510500052444}
}

@article{clauset2009power,
  title         = {Power-law distributions in empirical data},
  author        = {Clauset, Aaron and Shalizi, Cosma Rohilla and Newman, Mark EJ},
  journal       = {SIAM Review},
  volume        = {51},
  number        = {4},
  pages         = {661--703},
  year          = {2009},
  publisher     = {SIAM},
  doi           = {10.1137/070710111},
  url           = {https://doi.org/10.1137/070710111}
}

@book{mastatistical,
  title         = {Statistical Mechanics},
  author        = {Ma, SK},
  publisher     = {World Scientific, Singapore},
  year          = {1985},
  doi           = {10.1142/0073},
  url           = {https://doi.org/10.1142/0073}
}

@article{pakes69,
  title         = {Some conditions for ergodicity and recurrence of Markov chains},
  author        = {Pakes, AG},
  journal       = {Operations Research},
  volume        = {17},
  number        = {6},
  pages         = {1058--1061},
  year          = {1969},
  publisher     = {INFORMS},
  doi           = {10.1287/opre.17.6.1058},
  url           = {https://doi.org/10.1287/opre.17.6.1058}
}

@article{kingman61,
  title         = {The ergodic behaviour of random walks},
  author        = {Kingman, JFC},
  journal       = {Biometrika},
  pages         = {391--396},
  year          = {1961},
  publisher     = {JSTOR},
  doi           = {10.1093/biomet/48.3-4.391},
  url           = {https://doi.org/10.1093/biomet/48.3-4.391}
}

@article{thuraisingham2015a,
  title         = {Dementia and Hopfield model},
  author        = {Thuraisingham, RA},
  journal       = {Journal of Neural Transmission},
  volume        = {122},
  number        = {6},
  pages         = {773--777},
  year          = {2015},
  publisher     = {Springer},
  doi           = {10.1007/s00702-014-1339-3},
  url           = {https://doi.org/10.1007/s00702-014-1339-3}
}

@article{hu2015associative,
  title         = {Associative memory realized by a reconfigurable memristive Hopfield neural network},
  author        = {Hu, SG and Liu, Y and Liu, Z and Chen, TP and Wang, JJ and Yu, Q and Deng, LJ and Yin, Y and Hosaka, Sumio},
  journal       = {Nature Communications},
  volume        = {6},
  year          = {2015},
  publisher     = {Nature Publishing Group},
  doi           = {10.1038/ncomms8522},
  url           = {https://doi.org/10.1038/ncomms8522}
}

@article{peters2016evaluating,
  title         = {Evaluating gambles using dynamics},
  author        = {Peters, Ole and Gell-Mann, Murray},
  journal       = {Chaos: An Interdisciplinary Journal of Nonlinear Science},
  volume        = {26},
  number        = {2},
  pages         = {023103},
  year          = {2016},
  publisher     = {AIP Publishing},
  doi           = {10.1063/1.4940236},
  url           = {https://doi.org/10.1063/1.4940236}
}

@article{swendsen87,
  title         = {Nonuniversal critical dynamics in Monte Carlo simulations},
  author        = {Swendsen, Robert H and Wang, Jian-Sheng},
  journal       = {Physical Review Letters},
  volume        = {58},
  number        = {2},
  pages         = {86},
  year          = {1987},
  publisher     = {APS},
  url           = {https://doi.org/10.1103/PhysRevLett.58.86},
  doi           = {10.1103/PhysRevLett.58.86}
}

@book{davison97,
  title         = {{Bootstrap Methods and Their Application}},
  author        = {Davison, Anthony Christopher and Hinkley, David Victor},
  number        = {1},
  year          = {1997},
  publisher     = {Cambridge University Press},
  url           = {https://doi.org/10.1017/CBO9780511802843},
  doi           = {10.1017/CBO9780511802843}
}

@article{efron94,
  title         = {{An Introduction to the Bootstrap}},
  author        = {Tibshirani, Robert J and Efron, Bradley},
  volume        = {57},
  number        = {1},
  pages         = {1--436},
  year          = {1993},
  publisher     = {Chapman and Hall/CRC},
  url           = {https://doi.org/10.1201/9780429246593},
  doi           = {10.1201/9780429246593}
}

@dataset{suzen25zenV3,
  author        = {S\"{u}zen, Mehmet},
  title         = {Anomalous diffusion in convergence to effective ergodicity},
  month         = Dec,
  year          = 2025,
  publisher     = {Zenodo},
  doi           = {10.5281/zenodo.17936598},
  url           = {https://doi.org/10.5281/zenodo.17936598}
}

@article{sposini22,
  title         = {Towards a robust criterion of anomalous diffusion},
  author        = {Sposini, Vittoria and et. al.},
  journal       = {Communications Physics},
  volume        = {5},
  number        = {1},
  pages         = {305},
  year          = {2022},
  publisher     = {Nature Publishing Group UK London},
  url           = {https://doi.org/10.1038/s42005-022-01079-8},
  doi           = {10.1038/s42005-022-01079-8}
}

@article{munoz21,
  title         = {Objective comparison of methods to decode anomalous diffusion},
  author        = {Mu{\~n}oz-Gil, Gorka and et. al.},
  journal       = {Nature Communications},
  volume        = {12},
  number        = {1},
  pages         = {6253},
  year          = {2021},
  publisher     = {Nature Publishing Group UK London},
  url           = {https://doi.org/10.1038/s41467-021-26320-w},
  doi           = {10.1038/s41467-021-26320-w}
}

@article{cai25,
  title         = {Machine learning analysis of anomalous diffusion},
  author        = {Cai, Wenjie and Hu, Yi and Qu, Xiang and Zhao, Hui and Wang, Gongyi and Li, Jing and Huang, Zihan},
  journal       = {The European Physical Journal Plus},
  volume        = {140},
  number        = {3},
  pages         = {183},
  year          = {2025},
  publisher     = {Springer},
  url           = {https://doi.org/10.1140/epjp/s13360-025-06138-x},
  doi           = {10.1140/epjp/s13360-025-06138-x}
}

@article{firbas23,
  title         = {Characterization of anomalous diffusion through convolutional transformers},
  author        = {Firbas, Nicolas and Garibo-i-Orts, {\`O}scar and Garcia-March, Miguel {\'A}ngel and Conejero, J Alberto},
  journal       = {Journal of Physics A: Mathematical and Theoretical},
  volume        = {56},
  number        = {1},
  pages         = {014001},
  year          = {2023},
  publisher     = {IOP Publishing},
  url           = {https://doi.org/10.1088/1751-8121/acafb3},
  doi           = {10.1088/1751-8121/acafb3}
}

@article{seckler22,
  title         = {Bayesian deep learning for error estimation in the analysis of anomalous diffusion},
  author        = {Seckler, Henrik and Metzler, Ralf},
  journal       = {Nature Communications},
  volume        = {13},
  number        = {1},
  pages         = {6717},
  year          = {2022},
  publisher     = {Nature Publishing Group UK London},
  url           = {https://doi.org/10.1038/s41467-022-34305-6},
  doi           = {10.1038/s41467-022-34305-6}
}

@article{metzler98,
  title         = {Fractional diffusion, waiting-time distributions, and Cattaneo-type equations},
  author        = {Metzler, Ralf and Nonnenmacher, Theo F},
  journal       = {Physical Review E},
  volume        = {57},
  number        = {6},
  pages         = {6409},
  year          = {1998},
  publisher     = {APS},
  url           = {https://doi.org/10.1103/PhysRevE.57.6409},
  doi           = {10.1103/PhysRevE.57.6409}
}

@article{metzler00,
  title         = {The random walk's guide to anomalous diffusion: a fractional dynamics approach},
  author        = {Metzler, Ralf and Klafter, Joseph},
  journal       = {Physics Reports},
  volume        = {339},
  number        = {1},
  pages         = {1--77},
  year          = {2000},
  publisher     = {Elsevier},
  url           = {https://doi.org/10.1016/S0370-1573(00)00070-3},
  doi           = {10.1016/S0370-1573(00)00070-3}
}

@article{grossfield09,
  title         = {Quantifying uncertainty and sampling quality in biomolecular simulations},
  author        = {Grossfield, Alan and Zuckerman, Daniel M},
  journal       = {{Annual Reports in Computational Chemistry}},
  volume        = {5},
  pages         = {23--48},
  year          = {2009},
  publisher     = {Elsevier},
  url           = {https://doi.org/10.1016/S1574-1400(09)00502-7},
  doi           = {10.1016/S1574-1400(09)00502-7}
}

@article{wales10a,
  title         = {Energy Landscapes and Structure Prediction Using Basin-Hopping},
  author        = {Wales, David J},
  journal       = {Modern Methods of Crystal Structure Prediction},
  pages         = {29--54},
  year          = {2010},
  publisher     = {Wiley Online Library},
  url           = {https://doi.org/10.1002/9783527632831.ch2},
  doi           = {10.1002/9783527632831.ch2}
}

@article{tiampo02a,
  author        = {Tiampo, K. F. and Rundle, J. B. and McGinnis, S. and Gross, S. and Klein, W.},
  title         = {Mean-field threshold systems and phase dynamics: an application to earthquake fault systems},
  journal       = {Europhysics Letters},
  volume        = {60},
  number        = {3},
  pages         = {481--487},
  year          = {2002},
  doi           = {10.1209/epl/i2002-00289-y},
  url           = {https://doi.org/10.1209/epl/i2002-00289-y}
}

@article{tiampo03,
  author        = {Tiampo, K. F. and Rundle, J. B. and Klein, W. and S\'a Martins, J. S. and Ferguson, C. D.},
  title         = {Ergodic dynamics in a natural threshold system},
  journal       = {Physical Review Letters},
  volume        = {91},
  number        = {23},
  pages         = {238501},
  year          = {2003},
  doi           = {10.1103/PhysRevLett.91.238501},
  url           = {https://doi.org/10.1103/PhysRevLett.91.238501}
}

@article{tiampo07a,
  author        = {Tiampo, K. F. and Rundle, J. B. and Klein, W. and Holliday, J. and S\'a Martins, J. S. and Ferguson, C. D.},
  title         = {Ergodicity in natural earthquake fault networks},
  journal       = {Physical Review E},
  volume        = {75},
  pages         = {066107},
  year          = {2007},
  url           = {https://doi.org/10.1103/PhysRevE.75.066107},
  doi           = {10.1103/PhysRevE.75.066107}
}

@article{tiampo10a,
  author        = {Tiampo, K. F. and Klein, W. and Li, H. C. and Mignan, A. and Toya, Y. and Kohen-Kadosh, S. Z. L. and Rundle, J. B. and Chen, C. C.},
  title         = {Ergodicity and earthquake catalogs: Forecast testing and resulting implications},
  journal       = {Pure and Applied Geophysics},
  year          = {2010},
  doi           = {10.1007/s00024-010-0076-2},
  note          = {Published online},
  url           = {https://doi.org/10.1007/s00024-010-0076-2}
}

@article{cho10a,
  title         = {A simple metric to quantify seismicity clustering},
  author        = {Cho, NF and Tiampo, Kristy F and Mckinnon, SD and Vallejos, JA and Klein, W and Dominguez, R},
  journal       = {Nonlinear Processes in Geophysics},
  volume        = {17},
  number        = {4},
  pages         = {293--302},
  year          = {2010},
  publisher     = {Copernicus GmbH},
  url           = {https://doi.org/10.5194/npg-17-293-2010},
  doi           = {10.5194/npg-17-293-2010}
}

@article{li12a,
  title         = {Ergodicity examined by the Thirumalai-Mountain metric for Taiwanese seismicity},
  author        = {Li, Hsien-Chi and Chen, Chien-Chih},
  journal       = {Acta Geophysica},
  volume        = {60},
  number        = {3},
  pages         = {769--793},
  year          = {2012},
  publisher     = {Springer},
  url           = {https://doi.org/10.2478/s11600-012-0036-6},
  doi           = {10.2478/s11600-012-0036-6}
}

@article{baccetti24,
  title         = {Ergodicity, lack thereof, and the performance of reservoir computing with memristive networks},
  author        = {Baccetti, Valentina and Zhu, Ruomin and Kuncic, Zdenka and Caravelli, Francesco},
  journal       = {Nano Express},
  volume        = {5},
  number        = {1},
  pages         = {015021},
  year          = {2024},
  publisher     = {IOP Publishing},
  doi           = {10.1088/2632-959X/ad2999},
  url           = {https://dx.doi.org/10.1088/2632-959X/ad2999}
}

@article{barrows25,
  title         = {Resistive Ising: Effective resistance in random magnetic nanowire networks},
  author        = {Barrows, Frank and Iacocca, Ezio and Caravelli, Francesco},
  journal       = {Physical Review E},
  volume        = {112},
  number        = {1},
  pages         = {014302},
  year          = {2025},
  publisher     = {APS},
  url           = {https://doi.org/10.1103/hj7g-m5dc},
  doi           = {10.1103/hj7g-m5dc}
}

@article{saccone23,
  title         = {Real-space observation of ergodicity transitions in artificial spin ice},
  author        = {Saccone, Michael and Caravelli, Francesco and Hofhuis, Kevin and Dhuey, Scott and Scholl, Andreas and Nisoli, Cristiano and Farhan, Alan},
  journal       = {Nature Communications},
  volume        = {14},
  number        = {1},
  pages         = {5674},
  year          = {2023},
  publisher     = {Nature Publishing Group UK London},
  doi           = {10.1038/s41467-023-41235-4},
  url           = {https://doi.org/10.1038/s41467-023-41235-4}
}

@article{crater25a,
  title         = {Dipolar Aleppo lattice: Ground state ordering and ergodic dynamics in the absence of vertex frustration},
  author        = {Crater, Davis and Mahato, Gopi and Hoyt, Christian and Miertschin, Duncan and Regmi, Balaram and Hofhuis, Kevin and Dhuey, Scott and Achinuq, Barat and Caravelli, Francesco and Farhan, Alan},
  journal       = {Physical Review B},
  volume        = {111},
  number        = {14},
  pages         = {144407},
  year          = {2025},
  publisher     = {APS},
  url           = {https://doi.org/10.1103/PhysRevB.111.144407},
  doi           = {10.1103/PhysRevB.111.144407}
}

@article{melchert09,
  title         = {autoScale.py-A program for automatic finite-size scaling analyses: A user's guide},
  author        = {Melchert, Oliver},
  journal       = {arXiv preprint arXiv:0910.5403},
  year          = {2009},
  url           = {https://uol.de/f/5/inst/physik/ag/compphys/download/oliver/autoScale_guide.pdf}
}

@article{bhattacharjee01,
  title         = {A measure of datacollapse for scaling},
  author        = {Bhattacharjee, Somendra M and Seno, Flavio},
  journal       = {Journal of Physics A: Mathematical and General},
  volume        = {34},
  number        = {33},
  pages         = {6375},
  year          = {2001},
  publisher     = {IOP Publishing},
  url           = {https://doi.org/10.1088/0305-4470/34/33/302},
  doi           = {10.1088/0305-4470/34/33/302}
}

@inproceedings{qiu24,
  title         = {Scaling Collapse Reveals Universal Dynamics in Compute-Optimally Trained Neural Networks},
  author        = {Qiu, Shikai and Agarwala, Atish and Pennington, Jeffrey and Xiao, Lechao},
  booktitle     = {NeurIPS 2024, Workshop OPT 2024: Optimization for Machine Learning},
  url           = {https://opt-ml.org/papers/2024/paper78.pdf}
}

@book{risken,
  title         = {{Fokker-Planck Equation}},
  author        = {Risken, Hannes},
  pages         = {63--95},
  year          = {1989},
  publisher     = {Springer},
  url           = {https://doi.org/10.1007/978-3-642-61544-3_4},
  doi           = {10.1007/978-3-642-61544-3_4}
}

@manual{rlang,
  title         = {R: A Language and Environment for Statistical Computing},
  author        = {{R Core Team}},
  organization  = {R Foundation for Statistical Computing},
  address       = {Vienna, Austria},
  year          = {2021},
  url           = {https://www.R-project.org/}
}

@book{chambers08,
  title         = {Software for data analysis: programming with R},
  author        = {Chambers, John},
  year          = {2008},
  publisher     = {Springer},
  url           = {https://doi.org/10.1007/978-0-387-75936-4},
  doi           = {10.1007/978-0-387-75936-4}
}

@manual{isinglenzmc,
  title         = {isingLenzMC: Monte Carlo for Classical Ising Model},
  author        = {Mehmet S{\"u}zen},
  year          = {2014, 2025},
  note          = {{R package version 0.2.8}},
  doi           = {10.32614/CRAN.package.isingLenzMC},
  url           = {https://doi.org/10.32614/CRAN.package.isingLenzMC}
}

@book{privman, 
  title={{Finite Size Scaling and Numerical Simulation of Statistical Systems}},
  author={Privman, Vladimir},
  year={1990},
  publisher={World Scientific},
  url={https://doi.org/10.1142/1011},
  doi={10.1142/1011}
}

@article{nelder,
  title={Nelder-mead algorithm},
  author={Singer, Sa{\v{s}}a and Nelder, John},
  journal={Scholarpedia},
  volume={4},
  number={7},
  pages={2928},
  year={2009},
  url={http://www.scholarpedia.org/article/Nelder-Mead_algorithm}
}

@article{biroli24,
  title={Dynamical regimes of diffusion models},
  author={Biroli, Giulio and Bonnaire, Tony and De Bortoli, Valentin and M{\'e}zard, Marc},
  journal={Nature Communications},
  volume={15},
  number={1},
  pages={9957},
  year={2024},
  publisher={Nature Publishing Group UK London},
  url={https://doi.org/10.1038/s41467-024-54281-3},
  doi={10.1038/s41467-024-54281-3}
}

@article{serafino21,
  title={True scale-free networks hidden by finite size effects},
  author={Serafino, Matteo and Cimini, Giulio and Maritan, Amos and Rinaldo, Andrea and Suweis, Samir and Banavar, Jayanth R and Caldarelli, Guido},
  journal={Proceedings of the National Academy of Sciences},
  volume={118},
  number={2},
  pages={e2013825118},
  year={2021},
  publisher={National Academy of Sciences},
  url={https://doi.org/10.1073/pnas.2013825118},
  doi={10.1073/pnas.2013825118}
}

@article{fierro19,
  title = {Condensation of fluctuations in the Ising model: A transition without spontaneous symmetry breaking},
  author = {Fierro, Annalisa and Coniglio, Antonio and Zannetti, Marco},
  journal = {Phys. Rev. E},
  volume = {99},
  issue = {4},
  pages = {042122},
  numpages = {9},
  year = {2019},
  month = {Apr},
  publisher = {American Physical Society},
  doi = {10.1103/PhysRevE.99.042122},
  url = {https://link.aps.org/doi/10.1103/PhysRevE.99.042122}
}

@article{souza23,
  title = {Ergodicity and slow relaxation in the one-dimensional self-gravitating system},
  author = {Souza, L. F. and Filho, T. M. Rocha},
  journal = {Phys. Rev. E},
  volume = {107},
  issue = {1},
  pages = {014114},
  numpages = {9},
  year = {2023},
  month = {Jan},
  publisher = {American Physical Society},
  doi = {10.1103/PhysRevE.107.014114},
  url = {https://link.aps.org/doi/10.1103/PhysRevE.107.014114}
}

@article{sdobnov,
  title={Speckle dynamics under ergodicity breaking},
  author={Sdobnov, Anton and Bykov, Alexander and Molodij, Guillaume and Kalchenko, Vyacheslav and Jarvinen, Topias and Popov, Alexey and Kordas, Krisztian and Meglinski, Igor},
  journal={Journal of Physics D: Applied Physics},
  volume={51},
  number={15},
  pages={155401},
  year={2018},
  publisher={IOP Publishing},
  doi={10.1088/1361-6463/aab404},
  url={https://doi.org/10.1088/1361-6463/aab404}
}

\end{document}